\documentclass[conference, 9pt]{IEEEtran}
\IEEEoverridecommandlockouts
\IEEEpubid{\begin{minipage}{\textwidth}
    \ \\[45pt]  
    \footnotesize \hspace{-1.8cm}  
    \textcopyright~2025 IEEE. Personal use of this material is permitted. Permission from IEEE must be obtained for all other uses, in any current or future media, including reprinting/republishing this material for advertising or promotional purposes, creating new collective works, for resale or redistribution to servers or lists, or reuse of any copyrighted component of this work in other works. This work is accepted by the IEEE ASRU 2025.
\end{minipage}}
\usepackage{cite}
\usepackage{amsmath,amssymb,amsfonts}
\usepackage{algorithmic}
\usepackage{graphicx}
\usepackage{textcomp}
\usepackage{multirow}
\usepackage{booktabs}
\usepackage{hyperref}
\usepackage{cleveref}
\usepackage{float}
\usepackage{enumitem}
\usepackage{caption}
\usepackage{subcaption}
\usepackage{makecell}
\usepackage{dblfloatfix}
\usepackage{float}

\usepackage{xcolor}
\def\BibTeX{{\rm B\kern-.05em{\sc i\kern-.025em b}\kern-.08em
    T\kern-.1667em\lower.7ex\hbox{E}\kern-.125emX}}
\begin{document}

\title{Rapidly Adapting to New Voice Spoofing: Few-Shot Detection of Synthesized Speech Under Distribution Shifts}

\author{
    \IEEEauthorblockN{\textit{Ashi Garg, Zexin Cai, Henry Li Xinyuan, Leibny Paola Garc\'ia-Perera, Kevin Duh}\\ %
        \textit{Sanjeev Khudanpur, Matthew Wiesner, Nicholas Andrews}
        \vspace{.7\baselineskip}}
        \IEEEauthorblockA{\textit{Human Language Technology Center of Excellence, Johns Hopkins University}
}}
\maketitle

\begin{abstract}
We address the challenge of detecting synthesized speech under distribution shifts---arising from unseen synthesis methods, speakers, languages, or audio conditions---relative to the training data. Few-shot learning methods are a promising way to tackle distribution shifts by rapidly adapting on the basis of a few in-distribution samples. We propose a self-attentive prototypical network to enable more robust few-shot adaptation. To evaluate our approach, we systematically compare the performance of traditional zero-shot detectors and the proposed few-shot detectors, carefully controlling training conditions to introduce distribution shifts at evaluation time. In conditions where distribution shifts hamper the zero-shot performance, our proposed few-shot adaptation technique can quickly adapt using as few as 10 in-distribution samples---achieving upto 32\% relative EER reduction on deepfakes in Japanese language and 20\% relative reduction on ASVspoof 2021 Deepfake dataset.

\end{abstract}


\section{Introduction}
\label{sec:intro}
Recent advances in speech synthesis capabilities, including text-to-speech (TTS) and voice conversion (VC), have resulted in systems that can produce speech that is, in many cases, indistinguishable from actual human speech to a layperson~\cite{yang2024streamvcrealtimelowlatencyvoice, kameoka2018starganvcnonparallelmanytomanyvoice, choi2023dddmvcdecoupleddenoisingdiffusion, casanova2023yourttszeroshotmultispeakertts, qian2019autovczeroshotvoicestyle, kaneko2019cycleganvc2improvedcycleganbasednonparallel, bargum2023reimaginingspeechscopingreview}. 

However, such advancements in audio generation increase the risk of abuse. Watermarking techniques~\cite{cheng2001, faundez2006speaker, dathathri2024scalable} may be effective when applicable, but watermarking is not always a feasible defense and may be circumvented~\cite{liu2024audiomarkbench}. Therefore, it is critical to look into detection methods that make minimal assumptions about the synthesis methods.

Indeed, a number of supervised detectors have been proposed, which typically fit a neural network-based classifier to a dataset of real and fake utterances. 
While effective in certain settings, such detectors are vulnerable to real-world distribution shifts relative to the training data, such as the introduction of new speech synthesis methods, recording conditions, languages, and noise conditions~\cite{d2022underspecification,garg2025shiftyspeechlargescalesyntheticspeech}.

Since it is not possible to anticipate all possible distribution shifts and they are unavoidable in practice (i.e., outside of academic benchmarks), more adaptable detection approaches are needed. 

To this end, we remark that it is often feasible to collect a handful of presumed-fake speech in downstream applications, even if building a complete training set with thousands of samples is prohibitive. For example, a spoofing attempt could be detected using other information such as caller metadata. Alternatively, it is often possible to proactively synthesize small amounts of application-specific speech, as new synthesis methods are released, to simulate what an adversary may attempt.
This being the case, in this paper, we ask: based on a small sample of in-distribution speech---both synthetic and real---and a possibly larger set of out-of-distribution speech, can we build an accurate synthetic speech detector?

This problem is inherently difficult since it involves learning at test time, and there are risks of both overfitting and underfitting the small in-distribution sample.
We therefore turn to specialized methods designed for \emph{few-shot learning}.

Few-shot learning has been applied mainly for image classification tasks \cite{vinyals2017matchingnetworksshotlearning, finn2017modelagnosticmetalearningfastadaptation, snell2017prototypical}, and there is limited work in the speech domain (we discuss relevant work in~\autoref{sec:related_work}).

In this paper, we take a closer look at few-shot detection for synthetic speech detection. Specifically, we consider a broader set of experimental settings than previous work, and focus on settings where there is a controlled distribution shift from training-time to test-time. Furthermore, we consider state-of-the-art synthesis methods, evaluating across 12 different vocoders \cite{garg2025shiftyspeechlargescalesyntheticspeech} as well as multiple TTS and VC systems from the ASVspoof dataset \cite{wang2020asvspoof,10155166}. 

We demonstrate the importance of the feature aggregation strategy for few-shot learning by proposing a new self-attentive architecture for few-shot learning. 

In detail, our primary contributions are: 
\begin{itemize}[topsep=0pt,itemsep=-1ex,partopsep=1ex,parsep=1ex]
\item We perform a systematic study of few-shot learning for synthetic speech detection across a wide range of evaluation conditions, including controlled distribution shifts~\cite{garg2025shiftyspeechlargescalesyntheticspeech}.
\item We introduce a new few-shot learning method for synthetic speech detection---self-attentive prototypical networks---which significantly improves performance over state-of-the-art baselines in few-shot conditions.
\item We show that supervised fine-tuning of SSL-based detectors can exhibit strong performance in ``medium-shot'' settings involving larger adaptation sets.
\end{itemize}
To our knowledge, our work is the first to demonstrate the consistent benefit of few-shot learning for synthetic speech detection.

\section{Proposed method}
\subsection{Preliminaries}

\paragraph{Data} We assume access to two training datasets. The first is a potentially quite large set of real and synthesized speech, but is assumed to be out-of-distribution relative to the test conditions. We assume that real-or-fake labels are available for each speech sample and that fake data include speech generated from a variety of synthesis methods, for a diverse set of speakers. 
The first dataset is sufficiently large to train neural network-based detectors using gradient descent as baselines~(\autoref{sec:baselines}). The second dataset---the few-shot sample---comprises a small dataset of real and synthesized speech samples that is in-distribution relative to the test data. The challenge is to use this small sample to adapt to the test conditions, hopefully improving performance relative to models trained using only the first out-of-distribution dataset.

\paragraph{Pre-trained models} We leverage pre-trained speech representations $f_\theta$ obtained using self-supervised learning (SSL) models. These features have been shown to yield state-of-the-art zero-shot detectors\cite{tak2022automaticspeakerverificationspoofing} and therefore present a strong baseline for our proposed few-shot methods. Our approach aims to adapt these initial representations to novel test-time conditions. Although large-scale pre-training using SSL typically employs large datasets of real speech under diverse audio conditions, we assume that it is primarily capturing properties of genuine speech and any exposure to synthetic speech is incidental. However, SSL features are particularly well-suited in few-shot adaptation settings, since we hypothesize that they may be more readily fine-tuned using small samples of in-distribution data than representations trained from scratch, which is borne out in our experiments (see \autoref{tab:FT}).

\subsection{Attention-based Prototype Aggregation}

\paragraph{Prototypical Networks} We begin by introducing prototypical networks \cite{snell2017prototypical}---a simple but effective few-shot learning method which has seen various applications in image, text \cite{soto2024fewshotdetectionmachinegeneratedtext}, and audio \cite{heggan2022metaaudiofewshotaudioclassification} classification problems. The approach involves learning a mapping from a set of examples to a single vector \emph{prototype} of a class. For binary classification task of real and fake (spoof) speech detection, if we denote the prototype of the fake class $\mathbf{c}_{\text{fake}}$ and a learned few-shot embedding $g_\phi$, it is obtained via 

\begin{equation}\label{eqn:protonet}
\mathbf{c}_{\text{fake}} = \frac{1}{|S_\text{fake}|} \sum_{x \in S_{\text{fake}}} g_\phi (\mathbf{z}),
\end{equation}

\noindent where $\mathbf{z} = f_\theta(x)$ and $S_y$ is the support set for class $y$. Given a trained embedding and a test sample $x$, a classification is obtained via the softmax over the prototypes for each class (in our case, real-vs.-fake):

\begin{equation}
\label{eq:loss}
p(\texttt{fake} \mid x) = \frac{\exp(-d(g_\phi(\mathbf{z}), \mathbf{c}_{\text{fake}})}{\sum_{y \in \{ \texttt{real}, \texttt{fake} \} } \exp(-d(g_\phi(\mathbf{z}), \mathbf{c}_{y})}
\end{equation}

The parameters $\phi$ of embedding comprising the prototypical network are trained via episodic learning, where episodes of few-shot samples from each class are sampled repeatedly from a larger (out-of-distribution) training set. 

Here, the assumption is that there exists an embedding space where samples belonging to a particular class can be clustered around its class prototype. Further, the unlabeled query examples are classified based on their distance from the class prototypes. For the purpose of this work, we utilize Euclidean distance as the distance function $d(,)$ in Equation~\ref{eq:loss}.

\paragraph{Self-Attentive Prototypes} Note that the embedding function $g_\phi$ in~\autoref{eqn:protonet}, used in prototypical network, learns a representation of each data point independently. However, for synthetic speech detection, we hypothesize that the features to separate both classes may be more subtle and require a more expressive few-shot learning model. Inspired by  attentive pooling used in tasks like speaker verification~\cite{okabe18_interspeech}, we propose an attention-based aggregation method to compute more discriminative class prototypes. The new $g_\phi(x_1, x_2, \ldots)$ jointly embeds the entire support sample $x_1, x_2, \ldots \in S_y$, enabling learning higher-order features of the \emph{set} of examples rather than considering a simple mean. Specifically, given a set of support samples, to capture inter-sample dependencies and enhance each representation with global context, a multi-head self-attention (MHA) mechanism is applied over the support embeddings~\cite{vaswani2017attention}. We use a single layer of MHA with 2 attention heads. We then compute a weighted sum over these embeddings using a learnable attention mechanism to create the class prototype $\mathbf{c}_y$ as: 
\[
\mathbf{c}_y =\sum_{i=1}^N \alpha_i \tilde{\mathbf{z}_{i}}\]
for attention scores $\alpha$ and transformed outputs $\tilde{\mathbf{z}} = \text{MHA}(\mathbf{z})$, both parametrized by $\phi$. Finally, $\mathbf{c}_y$ is $\ell_2$-normalized. 

We note that our approach bears some resemblance to matching networks~\cite{vinyals2017matchingnetworksshotlearning}, which also consider ``full context embeddings,'' but in the case of matching networks, these are obtained using an LSTM which suffers from ordering biases---the ordering is arbitrary in the case of few-shot samples---and vanishing gradients which could be problematic with larger support sets.

\subsection{Binary vs Multi-class}
Synthetic speech detection is inherently a binary task at test time, where the goal is to distinguish between bonafide and spoofed audio. Nevertheless, we hypothesize that training with a multi-class classification objective, by treating each spoofing attack as a distinct class, could potentially help the model learn more fine-grained representations, thereby improving generalization to unseen attack types.

To investigate this, we examine the role of binary and multi-class classification training strategies in improving generalization to unseen spoofing attacks. For the multi-class setting, we use the attack labels provided in the ASVspoof 2019 training set, treating each attack type as a separate class. In the few-shot setting, we randomly sample three spoof classes per episode during training. For the binary setting, all spoof classes are merged into a single \textsc{fake} label to align with the test-time evaluation strategy.

\section{Experiments}
\label{sec:experiments}
In this section, we examine the adaptation ability of few-shot models to newly synthesized utterances. Our proposed models are compared against several baseline systems, including zero-shot spoof speech detection models and few-shot methods that utilize simple mean-pooling of speech representations.

\subsection{Model Implementation Details}
We utilize the state-of-the-art SSL-AASIST \cite{tak2022automaticspeakerverificationspoofing} model as the backbone architecture for developing the few-shot detector. SSL-AASIST integrates the Wav2Vec 2.0 XLSR \cite{conneau2020unsupervised} as a front-end feature extractor to obtain frame-level features with a spectro-temporal graph attention network as the back-end.  The spectral and temporal graphs are combined using two heterogeneous stacking graph attention layers (HSGAL) and two graph pooling layers for each of the branches. This layer helps take into account how artifacts from both spectral and temporal domains are related. 

\subsection{Baselines}\label{sec:baselines}
\paragraph{Anomaly detection} In non-stationary settings where test-time conditions are unknown, one effective approach is to concentrate modeling capacity on the known class—in our case, real speech. We then cast fake speech detection as an anomaly detection problem and identify deviations in feature space from the features of known real speech. Specifically, we employ the Mahalanobis distance-based approach described in~\cite{ren2021simple}, fitting a Gaussian to features extracted from in-distribution real speech samples. While this approach requires no samples of in-distribution fake speech, an important downside of the approach is that real speech that deviates from the training data may be falsely detected as fake.

\paragraph{Zero-shot} We include a number of state-of-the-art baseline zero-shot detectors, including SSL-AASIST and AASIST. These methods are often highly competitive, even in challenging settings like those considered in this paper, which involve distribution shifts between training and test time.

\paragraph{Few-shot} We compare against several ablations of our own method to validate the effectiveness of the proposed approach. Specifically, we consider the impact of varying the number of support samples $N$ for few-shot detection, the inclusion of the self-attentive pooling mechanism, and the choice of binary or multi-class classification objectives. We additionally include comparisons to~\cite{kukanov2024meta}, which report results on a subset of the datasets we consider, which we reproduce here.\footnote{At the time of this writing, code is not available to reproduce the results from~\cite{kukanov2024meta}.}

\paragraph{Supervised adaptation} We speculated in~\autoref{sec:intro} that the use of pre-trained SSL features may enable effective supervised adaptation using gradient descent, even with \emph{moderately} sized in-distribution fine-tuning sets of $N\geq100$ examples. We compare the performance of few-shot detectors with a fine-tuned SSL-AASIST model trained under the same data constraints. In the fine-tuning setup, we update the parameters of the zero-shot model using a small number of labeled samples per class through supervised learning. In contrast, the few-shot approach uses the same amount of data to compute class prototypes without updating the model weights. This allows us to evaluate whether prototype-based adaptation offers improved generalization with limited supervision. A key risk of the above baseline approach is overfitting to the small in-distribution sample. Therefore, we hold-out 30\% of the support sample for early stopping, and use the rest for supervised adaptation.

\subsection{Dataset}
We utilize ASVspoof 2019 dataset \cite{wang2020asvspoof} to train the detection systems. In order to assess generalization performance, we evaluate our models on four diverse and challenging datasets: ASVspoof 2021 \cite{10155166}, ShiftySpeech \cite{garg2025shiftyspeechlargescalesyntheticspeech}, In-the-Wild (ITW) \cite{müller2024doesaudiodeepfakedetection} and CodecFake \cite{wu2024codecfake}. These benchmarks collectively cover at least 10 distinctive evaluation conditions. ASVspoof 2021 extends the 2019 dataset with more advanced and varied spoofing attacks, including state-of-the-art TTS and VC systems. ShiftySpeech benchmark is specifically designed to evaluate model robustness under distributional shift, encompassing variations in vocoder type, language, and recording conditions. In contrast, the ITW dataset consists of deepfake samples sourced from various online platforms like YouTube, introducing uncontrolled acoustic variability and spoofing artifacts generated in the wild. Finally, CodecFake targets robustness against compression-induced artifacts by including spoofed speech compressed with diverse codecs and bitrates. Together, these datasets enable a comprehensive evaluation of generalization to previously unseen attacks and domain shifts.

\subsection{Experimental Setting}
The proposed few-shot models are trained with episodic learning \cite{snell2017prototypical} for 100 epochs, with each epoch comprising of 100 episodes. Each episode is a mini-batch designed to mimic the test-time scenario, where only a few examples from a novel class are available. 

During training, we set the number of support samples to create a prototype as 5 and the number of queries as 15 for each class. A similar setup is used for validation. The model is trained using prototypical loss with a learning rate of 1e-3 and optimized using the Adam optimizer. A StepLR learning rate scheduler is used, which decays the learning rate by a factor of 0.5 every 20 epochs.

We adopt the SSL-AASIST model as our backbone and, in our experiments, we freeze the Wav2Vec 2.0 XLSR-53 front-end while fine-tuning all remaining parameters. Model selection is based on accuracy measured on the validation set.

For zero-shot detectors, we utilize publicly available pre-trained checkpoints of SSL-AASIST\footnote{\url{https://github.com/TakHemlata/SSL_Anti-spoofing}} and AASIST\footnote{\url{https://github.com/clovaai/aasist}} models, both trained on ASVSpoof 2019 dataset. Additionally, we also train SSL-AASIST in a multi-class classification setting for 100 epochs, considering each attack type in ASVspoof 2019 as a distinct spoof class. This model is trained using a learning rate of 1e-5 and a batch size of 64. To investigate the effect of class granularity, we also train a multi-class variant of the few-shot model. In this setting, each episode samples two spoof classes randomly, with the bonafide class always included. The rest of the training setup remains consistent with the binary few-shot model.

For anomaly detection, we utilize x-vector \cite{snyder2018x} based speaker embedding model pretrained on VoxCeleb \cite{Chung18b} as the feature extractor\footnote{\url{https://huggingface.co/speechbrain/spkrec-xvect-voxceleb}. We also explored WavLM based speaker verification model as a feature extractor \url{https://huggingface.co/microsoft/wavlm-base-sv}}. We fit a multivariate Gaussian distribution on embeddings extracted from bonafide (real) samples in the ASVspoof 2019 dataset, estimating the mean vector \( \boldsymbol{\mu} \) and covariance matrix \( \boldsymbol{\Sigma} \). During inference, a given sample with embedding \( \mathbf{x} \) is scored based on the Mahalanobis distance.

The supervised adaptation detector with 10 samples per class is trained with batch size of 4, learning rate of 1e-4 and for 2 epochs.\footnote{To account for variability due to random sampling in 10-shot fine-tuning adaptation, we train each model independently 15 times per dataset} For 100 samples per class, the model is trained with a batch size of 64, learning rate of 1e-4 and for 3 epochs.

\vspace{2pt} \noindent \textbf{Evaluation Metric} To evaluate performance, we utilize Equal Error Rate (EER). EER is defined as the value where the false acceptance rate (FAR) equals the false rejection rate (FRR). To account for variability introduced by the random selection of support samples used to build prototypes, we evaluate the model over 100 runs with different randomly sampled support sets and report the mean EER. In zero-shot multi-class classification setting, we average the scores of all spoof classes to produce a single aggregated spoof score.

\subsection{Results and Discussion}

We evaluate zero-shot, few-shot, and fine-tuned models for synthetic speech detection across multiple datasets. The goal is to assess detection performance under limited supervision and distribution shift. Average Equal Error Rate (aEER) is used as the primary evaluation metric. Results are organized to reflect differences across domains and adaptation strategies.

\autoref{tab:aeer_comparison_shiftyspeech} reports aEER on ShiftySpeech dataset, while \autoref{tab:aeer_comparison_itw} reports results on in-domain (ASVspoof 2019) and more challenging out-of-domain datasets, including ASVspoof 2021 DF and In-the-Wild (ITW). 

\begin{table*}[hbt!]
\centering
\caption{Average EER (\% $\downarrow$) for zero-shot and few-shot methods on ShiftySpeech \cite{garg2025shiftyspeechlargescalesyntheticspeech}.}
\label{tab:aeer_comparison_shiftyspeech}
\begin{tabular}{@{}llcccccc@{}}
\toprule
\multirow{2}{*}{\textbf{Method}} & \multirow{2}{*}{\textbf{Setting}} & \multicolumn{6}{c}{\textbf{aEER (\%)}} \\
\cmidrule(lr){3-8}
& & \textbf{Ja} & \textbf{Zh} & \textbf{EN-Spk} & \textbf{EN-Audiobook} & \textbf{EN-Podcast} & \textbf{EN-YouTube}\\
\midrule
\multirow{3}{*}{Zero-shot} 
& AASIST &38.90 &67.61& 52.39& 49.49 &51.17 & 53.67\\
& SSL-AASIST & 22.15 & 26.25 & 32.40 & 34.58& 34.63 & 38.56 \\ 
& \makecell[l]{SSL-AASIST\\(multi-class)} &26.06 &29.93 &38.40&37.37& 39.71 &43.28\\
\midrule
\multirow{3}{*}{Few-shot ProtoNet (Binary)} 
&5-shot  &18.84 &31.12  & 32.52&30.19 & 37.55&37.80 \\
& 10-shot  & 18.26 & 29.48 & 30.67 & 28.40 & 35.58 & 36.13 \\
& 100-shot  & 17.98& 28.01 &29.24 &27.58&33.38& 33.62 \\
\midrule
\multirow{3}{*}{Few-shot ProtoNet + Attn Pool (Binary)} 
&5-shot  &15.53 &26.32  & 27.99&27.15 & 35.94&37.12 \\
& 10-shot  & 15.03 & 25.29 & 26.61 & 26.20 & 34.23 & 35.11 \\
& 100-shot  & \textbf{14.85}&\textbf{24.52}& \textbf{26.24}& \textbf{25.59} & \textbf{33.15}& \textbf{32.60} \\
\midrule
\multirow{3}{*}{Few-shot ProtoNet + Attn Pool (Multi-class)} 
&5-shot  &16.18 & 39.10 &32.44 & 30.78 & 37.35& 43.91 \\
& 10-shot  &15.69 & 36.01& 30.05& 28.77 &34.89 &39.67\\
& 100-shot  & 15.56& 31.29& 28.32 & 27.72 &32.65&34.04 \\
\bottomrule
\end{tabular}
\end{table*}

\begin{table*}[ht]
\centering
\caption{Average EER (\% $\downarrow$) for zero-shot and few-shot methods on in-domain and out-of-domain data (ITW, ASVspoof2021 DF).}
\label{tab:aeer_comparison_itw}
\begin{tabular}{llcccc}
\toprule
\multirow{2}{*}{\textbf{Method}} & \multirow{2}{*}{\textbf{Setting}} & \multicolumn{4}{c}{\textbf{aEER (\%)}} \\
\cmidrule(lr){3-6}
& & \textbf{ASV19:LA} & \textbf{ASV21:LA} & \textbf{ASV21:DF} & \textbf{ITW} \\
\midrule
\multirow{1}{*}{One-Class Learning} 
& Mahalanobis  & 19.42&24.93 & 43.26 &42.89 \\ 
\midrule
\multirow{3}{*}{Zero-shot} 
& AASIST &  0.99& 13.94 & 17.67 & 44.57\\
& SSL-AASIST &0.22 & 8.10& 8.33& 11.19 \\ 
& \makecell[l]{SSL-AASIST\\(multi-class)}  & 0.36& 11.41& 18.51& 13.45 \\
\midrule
\multirow{2}{*}{Few-shot\cite{kukanov2024meta}} 
& ProtoNet-256 & 1.68 & 4.55& 7.65 & 20.94 \\
& ProtoNet-96 & 1.68 & 4.56 & 7.65 & 21.04 \\
\midrule 
\multirow{3}{*}{Few-shot ProtoNet (SSL-AASIST, Binary)} 
&5-shot &2.66 &13.21 &6.65& 25.78 \\ 
&10-shot &2.43 &13.09&6.65&23.66\\ 
&100-shot &2.12 &12.95&6.65&23.36 \\
\midrule 
\multirow{3}{*}{Few-shot ProtoNet + Attn Pool (SSL-AASIST, Binary)} 
&5-shot & 6.24&11.52 &7.12 & 18.56\\ 
&10-shot &6.19&11.37 &6.95 & 18.56\\ 
&100-shot &6.26 &10.68 &6.52&18.56 \\
\midrule 
\multirow{3}{*}{Few-shot ProtoNet + Attn Pool (SSL-AASIST, Multi-class)} 
&5-shot & 0.72& 11.57&12.68& 21.77 \\ 
&10-shot & 0.62&10.67&11.13 &17.09\\ 
&100-shot & 0.58&10.26 &9.10&15.22 \\
\bottomrule
\end{tabular}
\end{table*}

\noindent \textbf{Zero-shot detection} As shown in \autoref{tab:aeer_comparison_shiftyspeech} and \autoref{tab:aeer_comparison_itw}, the SSL-AASIST model consistently outperforms the baseline AASIST model across all datasets. This highlights the natural benefit of incorporating self-supervised representations for spoof detection. For instance, on a challenging dataset such as YouTube, SSL-AASIST achieves a relative EER reduction of $\approx$ 28\% compared to AASIST. We further compare the generalization performance of zero-shot models with few-shot methods below.

\vspace{2pt} \noindent \textbf{Few-shot detection} Few-shot models demonstrate substantial performance gains over zero-shot baselines on average across the ShiftySpeech subsets. For instance, on Japanese (Ja) subset, the EER drops from 22.15\% (SSL-AASIST) using zero-shot model to 18.84\% with 5-shot learning. Similar trends are observed in the Audiobook and YouTube subsets, demonstrating that even limited supervision can yield substantial improvements. These gains are particularly pronounced under domain shift, where the spoofing artifacts differ significantly  from those seen during training on ASVspoof.

\begin{figure*}
    \centering
    \begin{subfigure}{.5\textwidth}
    \includegraphics[width=0.9\linewidth]{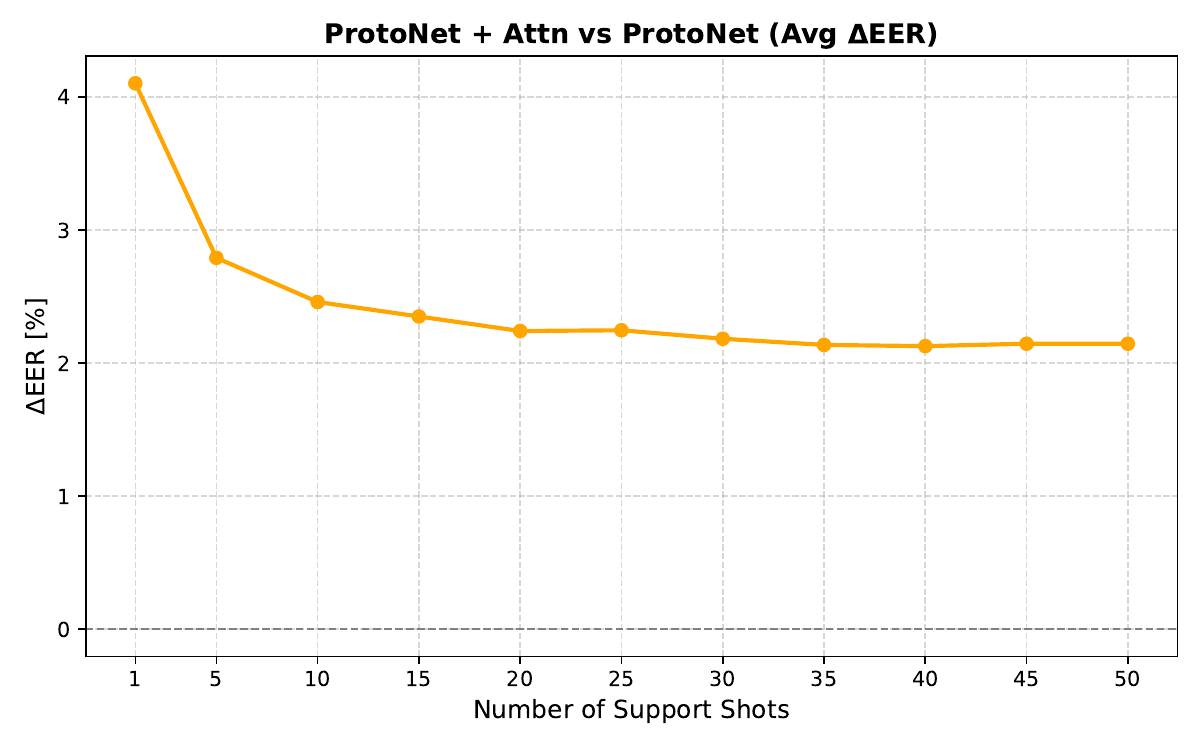}
    \caption{}
    \label{fig:eer_reduction}
    \end{subfigure}%
    \begin{subfigure}{.5\textwidth}
    \includegraphics[width=0.9\linewidth]
    {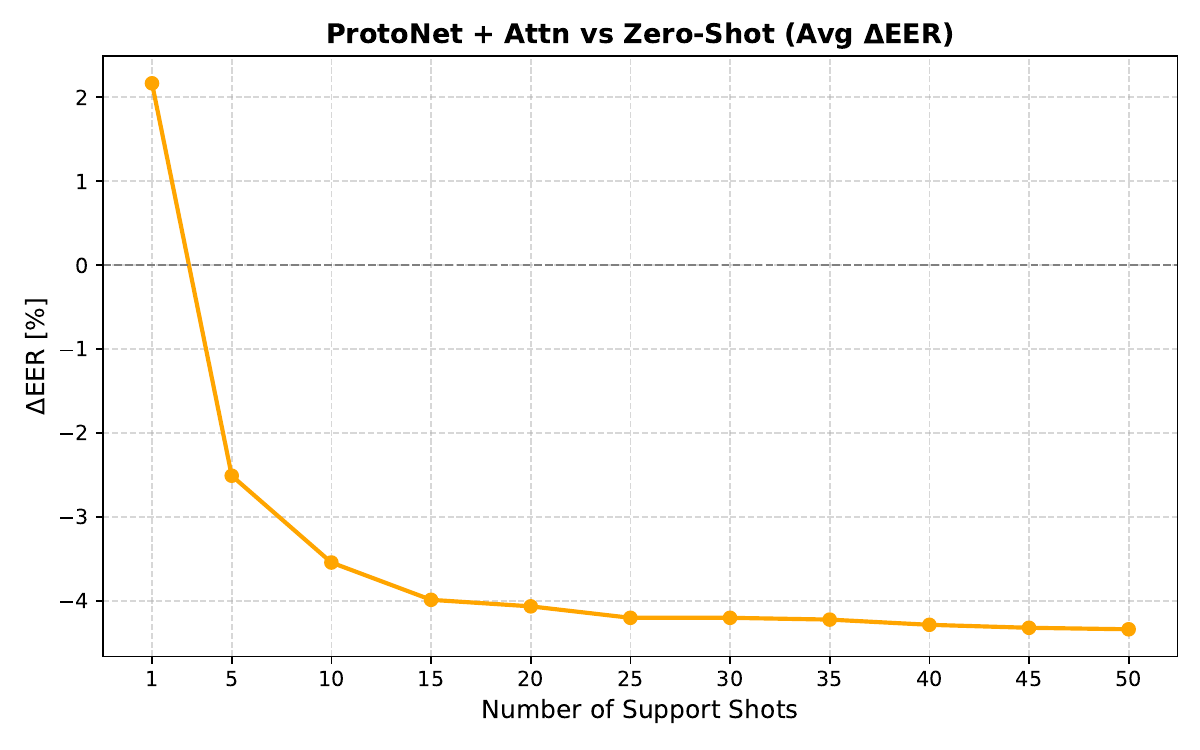}
    \caption{}
    \label{fig:eer_reduction_absolute}
    \end{subfigure}
    \caption{\textbf{Average $\Delta$EER} (ProtoNet $-$ ProtoNet+Attn) across datasets, computed per support shot $N$ (x-axis). \textbf{Left (a)}: A positive value indicates performance improvement using the proposed attention-based prototype aggregation over the ProtoNet baseline. \textbf{Right (b)}: a negative value indicates an \emph{improvement} in performance over the zero-shot baseline, which we see for $N \ge 5$.}
\end{figure*}

We also observe diminishing returns as the number of support samples increases from 10 to 100, indicating that only a small number of labeled examples are sufficient to achieve near-optimal results (see \autoref{fig:eer_reduction}). Specifically, increasing the number of support samples from 10 to 100 yields an $\approx$ 7\% relative EER reduction on the Podcast and YouTube datasets. However, there is less than 5\% relative reduction for other subsets of ShiftySpeech.

Additionally, incorporating attention pooling further enhances performance compared to standard mean-based prototype aggregation. For example, on Japanese subset, the 5-shot EER improves from 18.84\% (ProtoNet) to 15.53\% (ProtoNet + Attn Pool), and on Mandarin, from 31.12\% to 26.32\%---approximately a 15\% relative reduction in both cases.
These results suggest that attention-based prototype aggregation helps refine class representations, particularly under low-shot and cross-lingual conditions. For some challenging datasets such as ASVspoof DF, attention aggregation further improves performance over prior work \cite{kukanov2024meta}, utilizing over 200 samples. Specifically, attention aggregation yields 6.65\% EER on this set as compared to 7.65\% in prior work. Similar is observed for ITW, where 256 samples used in \cite{kukanov2024meta} yields an EER of 20.94\%, whereas utilizing only 5 support samples with an attention aggregation mechanism reduces the EER by 11.36\% relative to that baseline.

A similar trend is observed for codec-related shifts, as shown in the codec subsets (C1-C7) of the CodecFake dataset (\autoref{tab:codecfake_fewshot}). The zero-shot SSL-AASIST model achieves an average EER of 38.27\%. Notably, incorporating few-shot approach with as few as 5 support samples reduces the aEER to 31.98\%, highlighting strong generalization to previously unseen codec distortions.

\begin{table*}[ht]
\centering
\caption{Mean EER (\% $\downarrow$) on CodecFake C1–C7 subsets for different support set sizes.}
\begin{tabular}{llcccccccc}
\hline
\textbf{Method} & \textbf{Support} & \textbf{C1} & \textbf{C2} & \textbf{C3} & \textbf{C4} & \textbf{C5} & \textbf{C6} & \textbf{C7} & \textbf{aEER} \\
\hline
\multirow{1}{*}{SSL-AASIST } 
& Zero-shot      & 23.59 & 42.61 & 46.67 & 30.41 & 31.47 & 46.80 & 46.35 & 38.27 \\
\hline
\multirow{2}{*}{ProtoNet} 
& 5-shot &23.60 & 40.38 & 47.32 & 38.46 & 31.50 & 49.54 & 48.30 & 39.87  \\
& 10-shot  & 22.19 & 40.70 & 46.54 & 35.63 & 28.59 & 49.88 & 48.27 & 38.82 \\
& 100-shot & 19.98 & 37.17 & 43.70 & 34.03 & 27.58 & 48.53 & 45.85 & 36.69 \\
\hline
\multirow{2}{*}{\makecell[l]{ProtoNet + Attn Pool}} 
& 5-shot & 9.33 & 37.50 & 37.96  & 27.45 & 19.69 & 45.00 & 46.99 & 31.98\\
& 10-shot  & 9.27  & 35.53 & 34.88 & 26.03 & 19.69 & 43.58 & 45.07 & 30.57 \\
& 100-shot & 9.14  & 34.62 & 34.57 & 26.03 & 19.70 & 41.12 & 42.66 & \textbf{29.69} \\
\hline
\end{tabular}
\label{tab:codecfake_fewshot}
\end{table*}
 
In order to further quantify the performance gain from attention-based prototype aggregation, we compute the average difference in EER across datasets for each support set (See  \autoref{fig:eer_reduction}), defined as:
\begin{equation*}
    \Delta \text{EER}(k) = \frac{1}{N} \sum_{i=1}^N \left(\text{EER}_{\text{Proto}}(i,k) - \text{EER}_{\text{Proto+Attn}}(i,k) \right)
\end{equation*}
where $k$ is the number of support samples, and $N$ represents number of datasets.
In a similar manner, we also plot the performance gain over the best-performing zero-shot baseline (SSL-AASIST) and the few-shot detector trained with attention-based prototype aggregation. Particularly, we compute the average EER difference across all datasets for each support size $k$, defined as:
\begin{equation*}
\Delta \text{EER}(k) =  \frac{1}{N} \sum_{i=1}^N \left(\text{EER}_{\text{Proto+Attn}}(i,k) - \text{EER}_{\text{ZeroShot}}(i)\right)
\end{equation*}
A negative $\Delta \text{EER}(k)$ indicates that the few-shot model achieves a lower error rate than the zero-shot baseline. As shown in \autoref{fig:eer_reduction_absolute}, the performance improves consistently with increasing support size, but even small support size yield substantial gains highlighting the effectiveness of attention-based adaptation under limited supervision.

\vspace{2pt}\noindent \textbf{Binary vs Multi-class}
The zero-shot multi-class variant of SSL-AASIST improves over AASIST on most datasets; however, it performs worse than AASIST on ASVspoof 2021 DF (18.51 \% vs. 17.67 \%), and exhibits a notable degradation compared to the binary SSL-AASIST model (18.51 \% vs. 8.33 \%).

In the few-shot setting, the multi-class ProtoNet + Attn Pool (SSL-AASIST) model achieves strong performance on in-domain ASVspoof 2019 LA (0.72 \% with 5-shot) and competitive results on ASVspoof 2021 LA, but underperforms its binary counterpart in the more challenging out-of-domain datasets such as ASVspoof 2021, Zh and YouTube.
This trend is evident across datasets in both \autoref{tab:aeer_comparison_shiftyspeech} and \autoref{tab:aeer_comparison_itw}, suggesting that multi-class supervision may reduce generalization to unseen spoofing conditions.

\vspace{2pt} \noindent \textbf{Fine-tuning adaptation vs few-shot adaptation}
\autoref{tab:finetune_adapt} reports the EER on out-of-domain ShiftySpeech and ITW datasets. Across all datasets, few-shot adaptation consistently outperforms fine-tuning adaptation, when the number of samples per class is limited to 10. When the number of samples is increased to 100, fine-tuning model performance surpasses few-shot model. Notably, in the YouTube dataset, the fine-tuned model with 100 samples per class achieves an EER of 33.04\%, while the 10-shot few-shot model reaches 35.11\%, and the 100-shot few-shot model remains competitive at 32.60\%. These results indicate that few-shot learning provides a strong alternative to full-model fine-tuning under low-data conditions, while remaining competitive even when more data is available. However, we note that fine-tuning all model parameters at test-time may be impractical for large backbones ($f_\theta$) or when it is necessary to continually adapt to different test time conditions. Therefore, the 100-sample FT model may prove to be an upper bound on meta-learning approaches, which have the ostensibly harder job of adapting without changing the underlying model parameters.

\begin{table*}[htb!]
\centering
\caption[]{Comparison of model fine-tuning (FT) adaptation and few-shot adaptation under few-shot ($N=10$) and medium-shot ($N=100$) conditions. For $N=100$, the pre-trained SSL features enabled quick adaptation to the test conditions without significant overfitting of the small training sample. However, in the few-shot setting ($N=10$), this approach breaks down.}
\label{tab:finetune_adapt}
\begin{tabular}{lccccccc}
\toprule
\textbf{Model} & \textbf{Ja} & \textbf{Zh} & \textbf{EN-Spk} & \textbf{EN-Audiobook} & \textbf{EN-Podcast} & \textbf{EN-YouTube} & \textbf{ITW} \\
\midrule
SSL-AASIST (FT, 10-sample)& 26.60 & 41.53 & 39.31& 45.97& 45.69 &46.19 &  24.34\\
SSL-AASIST (FT, 100-sample) &4.33 & 13.58 & 13.31 & 17.61& 24.18 & 33.04 & 4.71 \\
Few-shot (10-shot) &15.03& 25.29 & 26.61 & 26.20& 34.23& 35.11 & 18.56 \\
Few-shot (100-shot) &14.84 & 24.52& 26.24& 25.59 & 33.15& 32.60  & 18.56\\
\bottomrule
\end{tabular}
\label{tab:FT}
\end{table*}

\noindent \textbf{Anomaly detection} We evaluate the one-class learning approach on both in-domain and the more challenging ITW dataset. This method yields a relatively high EER of 19.42\% even on the in-domain setting, and an EER of 42.89\% on ITW. Notably, the EER on ITW is comparable to that of the zero-shot AASIST model (44.57\%). Although performance is limited compared to supervised methods, these result suggest the potential of utilizing only bonafide samples for detection. 

\section{Related Work}\label{sec:related_work}
Prior work in deepfake detection has been driven by a number of key benchmarks \cite{xiao2015spoofing,wang2020asvspoof,10155166}. However, for detection research to translate into real-world effectiveness, cross-domain generalization is crucial for developing detectors that are robust to real-world distribution shifts. Few-shot learning has been employed to enhance performance in low-resource settings and to improve generalization to unseen classes or domains. Such methods have found applications in variety of audio classification tasks, including speaker identification \cite{wolters2020study,chen2018investigation,liang2025self}. However, its application to synthetic speech detection remains underexplored. Only a few recent studies have investigated meta-learning approaches \cite{kukanov2024meta, pal2022synthetic}. \cite{kukanov2024meta} focuses on utilizing various loss functions to improve generalization but limits evaluation to the ASVspoof dataset. \cite{pal2022synthetic} further investigates by utilizing self-supervised features to study the generalization performance of vanilla Prototypical Networks on more challenging datasets such as ITW. 
In this work, we take a closer look at few-shot adaptation for synthetic speech detection, and propose an improvement over standard prototypical networks which incorporates attention-based prototype aggregation.

\section{Conclusion}\label{sec:conclusion}

\noindent \textbf{Summary} We have shown that few-shot methods are able to significantly improve detection performance relative to state-of-the-art zero-shot detectors, given only a small in-distribution samples ($N=5$ or $N=10$). 
Specifically, we find that learning set-level embeddings using self-attention is a key ingredient in unlocking the potential of few-shot detection in challenging test conditions. Overall, in the few-shot setting, the proposed few-shot adaptation approach greatly outperforms alternatives based on supervised fine-tuning or anomaly detection as demonstrated using datasets like ShiftySpeech \cite{garg2025shiftyspeechlargescalesyntheticspeech}, which facilitated systematic evaluation under a wide range of distribution shifts.. For larger numbers of samples ($N=100$), we find that full fine-tuning of SSL-based detectors obtains strong performance, albeit at much larger computational cost and by producing specialized models.

\vspace{2pt} \noindent \textbf{Limitations \& Future Work} Our work focuses on a single SSL backbone architecture (SSL-AASIST). While this enabled controlled comparisons between zero-shot and few-shot learning methods, it is possible that other pre-trained speech representations could improve our results further. Exploring other meta-learning strategies, such as Model-Agnostic Meta-Learning (MAML)\cite{finn2017model} and model architectures, may yield further improvement; however, these meta-learning methods incur a much greater computation burden, making them difficult to evaluate across all the conditions we consider here. 

Additionally, our work focuses on features derived from neural networks; however, low-level features such as pitch, subband features, spectral features, and harmonic-to-noise ratio could also be discriminative and useful in few-shot settings \cite{yang2019significance,7882691,xiao2015spoofing,patel2015combining}.

\section*{Acknowledgment}
This work was supported by the Office of the Director of National Intelligence (ODNI), Intelligence Advanced Research Projects Activity (IARPA), via the ARTS Program under contract D2023-2308110001. The views and conclusions contained herein are those of the authors and should not be interpreted as necessarily representing the official policies, either expressed or implied, of ODNI, IARPA, or the U.S. Government. The U.S. Government is authorized to reproduce and distribute reprints for governmental purposes notwithstanding any copyright annotation therein.

\newpage
\bibliographystyle{IEEEtran}
\bibliography{ref}

\vspace{12pt}

\end{document}